\documentstyle[aps,prl,epsf,multicol]{revtex}

\begin{document}
\draft
\title{Magnetotunneling as a Probe of Luttinger-Liquid Behavior}

\author{Alexander Altland$^{1,2}$, C. H. W. Barnes$^2$, 
F. W. J. Hekking$^{1,2,3}$ 
and A. J. Schofield$^{2,4}$}
\address{
$^1$ Theoretische Physik III, Ruhr-Universit\"at Bochum, 44780 Bochum,
Germany\\
$^2$ Cavendish Laboratory, University of Cambridge, Madingley Road, 
Cambridge CB3 OHE, United Kingdom\\
$^3$ CNRS-CRTBT \& Universit\'e Joseph Fourier, 38042 Grenoble Cedex 9, France\\
$^4$ School of Physics and Astronomy, University of Birmingham, Birmingham B15 2TT, United
Kingdom\\
}

\date{\today}
\maketitle 
\begin{abstract}
A novel method for detecting Luttinger-liquid behavior is
proposed. The idea is to measure the tunneling conductance between a
quantum wire and a parallel two-dimensional electron system as a function
of both the potential difference between them, $V$, and an in-plane
magnetic field, $B$. We show that the two-parameter dependence on $B$
and $V$ allows for a determination of the characteristic dependence on
wave vector $q$ and frequency $\omega$ of the {\it spectral function},
$A_{\rm LL}(q,\omega)$, of the quantum wire. In particular, the
separation of spin and charge in the Luttinger liquid should manifest
itself as singularities in the $I$-$V$-characteristic. The
experimental feasibility of the proposal is discussed.
\end{abstract} 

\begin{multicols}{2}
\narrowtext 

The physical properties of a one-dimensional electron system (1DES)
are markedly distinct from those of its higher dimensional
counterparts: No matter how weak the interactions
between particles, the 1DES cannot be described within established
Fermi-liquid like pictures of interacting fermions. Rather, it is
always unstable towards the formation of a highly correlated state of
matter, the so-called Luttinger liquid (LL)~\cite{haldane81}.
LL-behavior is signalled by the absence of electron-like
quasiparticles and instead is characterized by separate low-lying
collective excitations associated with spin and charge degrees of
freedom. This phenomenon of spin-charge separation and other features
identifying LL phases have been studied extensively and various
excellent reviews on the subject
exist~\cite{emery79,solyom79,schulz95,voit95,fisher96}.  The continued
research activity on one-dimensional systems is not merely of
academic interest as there are a growing number of physical
applications: organic polymers~\cite{heeger88}; carbon
nanotubes~\cite{thess96}; quantum Hall edge
states~\cite{wen90,milliken96}; and ultra-narrow quantum
wires~\cite{tarucha95yacoby96} are believed to fall into the general
class of 1DES's.
  
Despite this the present experimental situation is inconclusive.
Although previous studies on organic conductors and superconductors,
inorganic charge density wave materials, semiconductor quantum wires
and fractional quantum Hall phases (see~\cite{voit95} for a more
extensive list of references) have been {\it consistent} with various
aspects of the highly correlated behavior of 1DES's, an unambiguous
experimental observation of a LL-phase is still lacking.

\begin{figure}
\centerline{\epsfxsize=3.5in\epsfbox{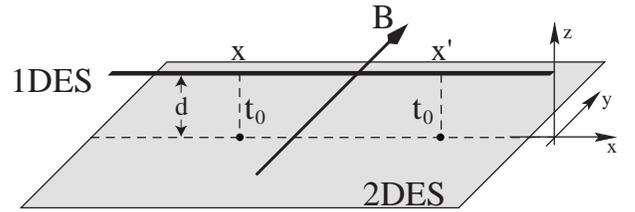}}
\vspace{0.5cm}
\caption{\label{fig:1}Device configuration for magnetotunneling between
a 1DES and a 2DES.}
\end{figure}

In this Letter we propose a novel experiment---falling into the
general class of semiconductor transport measurements---which should
provide evidence for spin-charge separation in 1D.  The basic
experimental device is displayed in Fig.~\ref{fig:1}.  A 1DES runs
at a height $d$ above a parallel two-dimensional electron system
(2DES).  The 1DES and the 2DES are kept at a relative voltage $V$ and
an in-plane magnetic field is applied with a component $B$
perpendicular to the wire. A setup of this type  may be
realized in a number of ways: a double quantum well (DQW)
heterostructure patterned with appropriate external gates
\cite{kardynal97}; a suitably etched resonant tunneling diode
\cite{Beton}; or an organic polymer or carbon nanotube with an
electrical contact at one end \cite{heeger88,thess96} 
placed on an undoped heterostructure with a shallow 2DES. 
The presence of a voltage bias induces the flow of a tunnel current
$I(V,B)$ between the 1DES and the 2DES. As will be detailed below,
$I(V,B)$ is essentially determined by the overlap of the spectral
functions $A_i$, $(i=2D,1D)$ of the two subsystems.  By
fine-tuning the control parameters $V$ and $B$ the overlap integral
changes in a pronounced way, thereby probing features of both $A_{1D}$
and the (essentially known) $A_{2D}$. The former is believed to be
governed by the phenomenon of spin-charge separation.  In this way,
the 2DES can be employed as a `spectrometer' scanning the LL
characteristics of the quantum wire. 
To formulate the above program quantitatively we model  
the device depicted in Fig.~\ref{fig:1} in terms of the Hamiltonian
\begin{equation}
\label{ham}
H = H_{1D} + H_{2D} + H_{T},
\end{equation}
where $H_{1D}$, $H_{2D}$ describe the 1DES and the 2DES, respectively.
The tunnel Hamiltonian $H_{T}$ transfers electrons between the 1DES and the
2DES. It is modeled as
\begin{equation}
H_{T}
=
t_0\int dx 
[ e^{-iedBx}
\Psi ^{\dagger}_{2D,s} (x)
\Psi _{1D,s}^{\vphantom{\dagger}} (x)
+
\mbox{h.c.}
] ,
\label{tunham}
\end{equation}
where $\Psi _{i,s}, i=1D,2D$ are fermionic field operators with spin $s=
\uparrow, \downarrow$ and $\Psi_{2D}(x)$ is a shorthand for the 2DES field
operator evaluated at point ${\bf x}=(x,0)$ (see Fig.~\ref{fig:1}).  We have
chosen a gauge where the entire dependence on the magnetic field is contained
in the Aharonov-Bohm phases carried by the matrix elements of $H_T$. [In
passing we note that the magnetic field needed to drive the effects discussed
below is weak: $B\sim eV/(dv_F)$, where $v_F$ is the Fermi velocity of the
2DES.  Fields of this type are not expected to affect the bulk physics of both
the 1DES and the 2DES.]  In writing (\ref{ham}) and (\ref{tunham}) two
essential approximations have been made: First, drag effects (i.e.
electron-electron interactions between 1DES and 2DES) are neglected. The
justification is that at the low temperatures considered here, standard Fermi
liquid arguments applied to the 2DES show that drag effects are suppressed by
a phase space factor $\sim T^{2}$ at temperature $T$.  Second, it is assumed
that tunneling occurs between neighboring points $x\in {\,\rm
  1DES\,}\leftrightarrow {\bf x} \in {\, \rm 2DES}$ only (with amplitude
$t_{0}$). Owing to the exponential dependence of the tunneling amplitude on
both the height of the tunneling barrier and the tunneling distance, direct
processes are the most relevant by far.  By virtue of this assumption, the
problem becomes effectively one-dimensional.

To leading order in the amplitude $t_0$ the tunnel current per unit
length is given by~\cite{mahan}:
\begin{eqnarray}
&&I(V,B)
=
\frac{4I_0}{m}\int dq \int
\frac{d\epsilon}{2\pi}
[f(\epsilon -eV) - f(\epsilon )]\nonumber\\
&&\qquad\times A_{1D}(q,\epsilon )
A_{2D}(q-q_{B},\epsilon -eV),
\label{tuncur}
\end{eqnarray}
where $f(\epsilon)$ is the Fermi function, $m$ the 2DES electron mass
and $I_0=e|t_0|^2 m/\pi$ the natural unit of current in the
problem. The spectral functions $A_i(q,\omega) = -2 \Im
\mbox{m } G_i^R (q,\omega)$, where
$G_i^R(q,\omega)$ are the Fourier transforms of the retarded Green
functions $G_i^R(x,t) = -i \theta (t) \langle \{\Psi _{i,s} (x,t) ,
\Psi _{i,s}^\dagger (0,0)\} \rangle$~\cite{fn2}.  

The structure of the above integral representation of $I(V,B)$ already
reveals the basic idea of this Letter: according to (\ref{tuncur}) the
current is given by the overlap of the two spectral functions
integrated over a window of width $\max(T,eV)$ at the Fermi energy.
As detailed below, the value of the overlap integral sensitively
depends on the two parameters $eV$ and $q_B\equiv eBd +
k_{F}^{2D}-k_{F}^{1D}$ which shift the relative origin of the two
spectral functions. $k_{F}^{1D/2D}$ are the Fermi wave vectors of the
1DES and 2DES respectively.  The 1DES $A_{1D}(q,\omega)$ is expected
to exhibit pronounced structures depending in a non-trivial way on LL
characteristics, whereas the spectral function of the 2DES is
dominated by electron-like quasiparticles and its important features
are explicitly known.  Thus, $A_{2D}$ may serve as a `spectrometer'
scanning the features of $A_{1D}$ as $q_B$ and $eV$ are varied. In
particular, {\it assuming} that $A_{1D}$ is of LL type we show below
that the tunnel current is profoundly affected by the phenomenon of
spin-charge separation which should give a clear
signal of LL behavior.

\begin{figure}
\centerline{\epsfxsize=3.3in\epsfbox{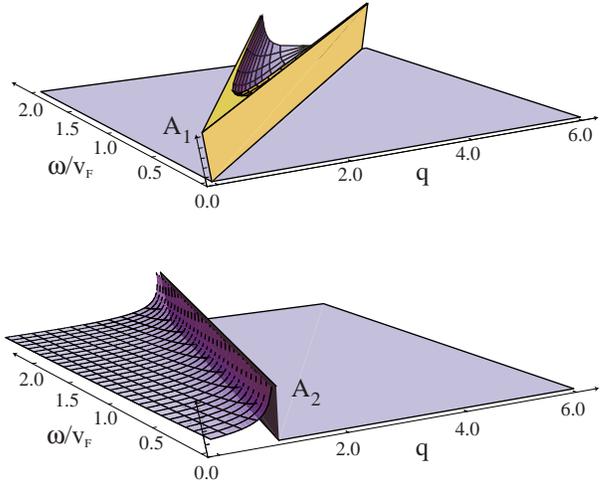}}
\vspace{0.5cm}
\caption{\label{fig:2}Plot of the two spectral functions in the
  $(q,\omega)$ plane (arbitrary units) for $v_\rho=v_F/2$, $v_\sigma=v_F/3$
  and $r=1.5$.}
\end{figure}

\begin{figure}
\centerline{\epsfxsize=3.5in\epsfbox{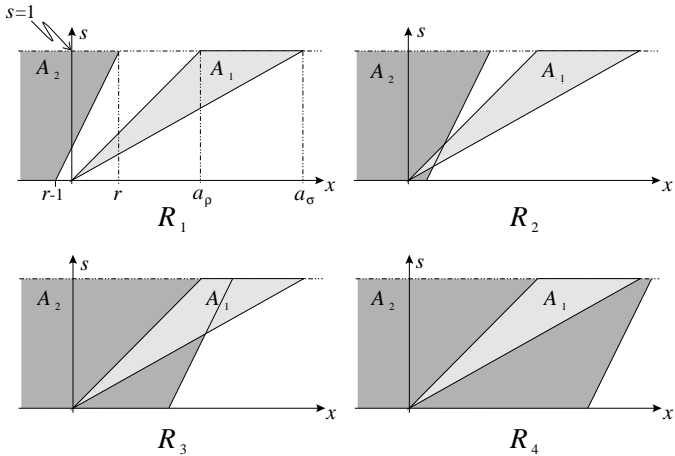}}
\vspace{0.5cm}
\caption{\label{fig:3}Relative position of the two spectral functions
  in the four regimes $R_1,\dots,R_4$. The light (dark) shaded areas
  represent the functions $A_{1D}$ $(A_{2D})$.}
\end{figure}

We proceed by specifying the spectral functions employed in
calculating the current. Owing to the one-dimensionality of the problem
both functions $A_i$ can be decomposed according to $A_i(q) =
\sum_{\eta=\pm1} A_{i,\eta}(q_\eta,\omega)$, where $A_{i,1}$ ($A_{i,-1}$)
represents the contribution of right- and left-moving charge carriers,
respectively.  Assuming that both interactions and disorder are
negligible (an assumption we discuss below), the
function $A_{2D}$ in the vicinity of the Fermi surface
is then given by (see Fig.~\ref{fig:2})
\begin{equation}
A_{2D,\eta}(q,\omega)
=\sqrt{2m}
\frac{\Theta(\omega -\eta qv_{F})}{\sqrt{\omega -\eta qv_{F}}}.
\label{A2D}
\end{equation}
As for $A_{1D}$, various forms of LL
spectral functions have been discussed in the literature. We here
employ the function (see Fig.~\ref{fig:2})
\begin{eqnarray}
\label{ALL}
&&A_{1D,\eta}(q,\epsilon)
=2\\
&&\frac{\Theta(\epsilon -\eta qv_{\sigma})
\Theta(\eta qv_{\rho}-\epsilon)
      +
      \Theta(\eta qv_{\sigma}-\epsilon )
      \Theta(\epsilon -\eta qv_{\rho})}
      {\sqrt{|\epsilon  -\eta qv_{\rho}||
      \epsilon -\eta qv_{\sigma}|}},\nonumber
\end{eqnarray}
where $v_\sigma$ and $v_\rho$ are the velocities of spin and charge
density waves, respectively. For the type of systems considered here,
$v_F > v_\rho > v_\sigma$\cite{fn1}. Eq.(\ref{ALL}) was derived in
Ref.~\cite{voit95} under the simplifying assumption of no interactions
between left and right moving particles (formally that the
Luttinger-Liquid parameter $K_\rho=1$). This condition can be relaxed
at the expense of the appearance of spectral weight outside the limits
defined by (\ref{ALL}).  This does not alter the main conclusions of
this Letter but would add considerably to the complexity of exposition.
We therefore leave it for subsequent discussion \cite{altland98}.
Substituting Eqs.~(\ref{A2D}) and (\ref{ALL}) into (\ref{tuncur}), we
find that four regimes $R_j, j=1,\dots,4$ with qualitatively different
behavior exist. Introducing dimensionless parameters $r =
1+q_{B}v_{F}/eV$, $a_\rho = v_F/v_\rho$ 
and $a_\sigma = v_F/v_\sigma$, these
are given by  $R_1:r<1, R_2:1\le r \le a_\rho,
R_3: a_\rho \le r < a_\sigma$ and $R_4: r>a_\sigma$. A schematic plot
of the relative positioning of the spectral functions in the regimes
$R_1,\dots,R_4$, respectively, is shown in Fig.~\ref{fig:3} as a 
function of the dimensionless 1DES wave vector $x=qv_{F}/eV$ and
frequency $s=\omega/eV$.

The current can now be obtained by double
integration over $q$ and $\omega$. In all regimes
the integrations can be carried out in closed form although the
resulting formulae tend to be somewhat lengthy and partly involve
special functions so will be discussed elsewhere~\cite{altland98}. Here we
restrict ourselves to a discussion of the current in the asymptotic
regions where adjacent regimes meet (and the sensitivity of the result
to variations in the external parameters is most pronounced).

\begin{figure}
\centerline{\epsfxsize=3in\epsfysize=3in\epsfbox{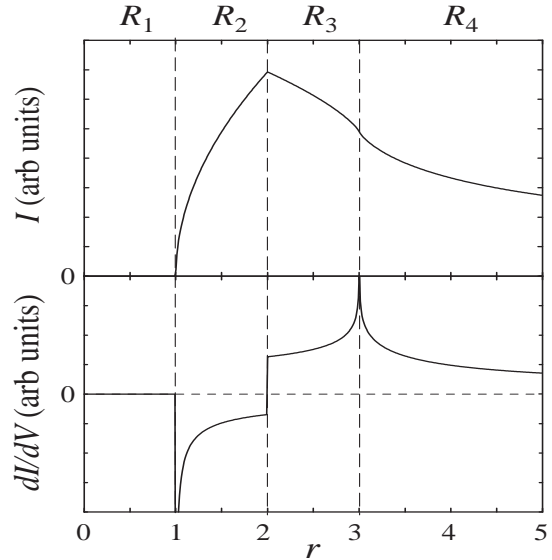}}
\vspace{0.5cm}
\caption{\label{fig:4}
Tunneling current and differential conductance as a
function of the magnetic field ($a_\rho=2$, $a_\sigma=3$).}
\end{figure}

Figure~\ref{fig:4} shows both $I$ and the differential conductance $G\equiv
dI/dV$ at $T=0$ plotted as a function of the parameter $r$.  We note here that
variation of $r$ may be achieved not only by varying $B$ but also by changing
the relative 2DES or 1DES carrier densities.  However, this would introduce
the possibility for capacitive coupling effects which would make the
determination of the LL parameters more difficult~\cite{altland98}.  In the
following we discuss the behavior of the result in the various regimes
separately.

{\it $R_1$:} The two spectral functions $A_{1D}$ and $A_{2D}$ do
not overlap (cf. Fig.~\ref{fig:3}) implying that the current
vanishes.

{\it $R_2$:} For  $r>1$, the spectral functions start to overlap
leading to a (singular) onset of current flow. At the same time the
conductance diverges as $\tilde g\sim -(r-1)^{-1/2}$, where we have 
introduced $\tilde g \equiv G \sqrt{ Ve^{-1}E_F}/I_0)$ as a
dimensionless measure for the conductance. The inverse square root
behavior of the conductance persists 
up to the boundary to $R_3$ where
$g(r\to a_\rho) = -(a_\rho -1 )^{-1} (a_\rho a_\sigma
(a_\sigma-1))^{1/2}$. 

{\it $R_3$:} As $r$ crosses over into $R_3$, the conductance exhibits
a second discontinuity, the magnitude of which is found to be
\[
\textstyle
g(a_\rho^+) - g (a_\rho^-) = 
\frac{a_\rho}{a_\rho-1}\sqrt{\frac{a_\rho a_\sigma}{a_\sigma -
    a_\rho}},
\]
where $a_\rho^\pm = a_\rho \pm \delta$, $\delta$ infinitesimal. Note
that the jump is accompanied by a change of sign. As $r$ approaches
the boundary to $R_4$, the conductance again exhibits a singularity,
this time of logarithmic type. More precisely,
\[
\textstyle
g(r) \stackrel{r\to a_\sigma}\longrightarrow
-\frac{a_\sigma}{a_\sigma-1} \sqrt{\frac{a_\rho a_\sigma}{a_\sigma -
    a_\rho}} \frac{1}{\pi}\ln(a_\sigma-r).
\]

{\it $R_4$:} The boundary singularity at $a_\rho$ turns out to be
symmetric, i.e. for small $\epsilon$, $g(r=a^\sigma+\epsilon)=
g(r=a^\sigma-\epsilon)$. Eventually, for asymptotically large $r$ the
conductance decays as $g \sim r^{-1/2}$.

In summary we see that the structure of the $I-V$ characteristic is
essentially determined by the two spin-charge parameters $a_\rho$ and
$a_\sigma$.  For a more general Luttinger-liquid, with $K_\rho\neq 1$,
the power laws associated with the singular features will be modified
but their location is determined by $a_\rho$ and $a_\sigma$ allowing
these parameters to be measured.  In order to decide whether this
strategy of demonstrating LL behavior is practical
it is imperative to estimate the effect of two ingredients that tend
to blur the above sharp structures of the $I-V$-curve: finite
temperatures and disorder.

As for the effect of finite temperatures, it is intuitively clear that the
structures of the $I-V$ characteristics will be completely smeared for 
$T$ larger than any of the characteristic energy scales ($eV
a_{\rho,\sigma}, q_B v_F a_{\rho ,\sigma}$ or any combination thereof)
of the problem. (To see this more explicitly, notice that
for finite $T$ the integration in (\ref{tuncur}) no longer extends
over a sharply defined strip in the $(q,\omega)$-plane but rather over
a smeared region of width $eV\pm T$). However, it has been 
demonstrated for 2D-2D and 1D-2D tunneling in DQW 
structures that at temperatures readily available in experiment
its effect may be ignored \cite{kardynal97,Eisen,Neil}.

The effects of disorder are more significant and can be considered
individually for the 2DES and 1DES.  For the 2DES, 2D-2D tunneling
measurements show an effective blurring of $A_{2D}$ over an energy
range $\Gamma$, where $\tau=\Gamma^{-1}$ is the average scattering
time in the 2DES.  Optimizing $\Gamma$ to be smaller than the
characteristic energy scales of the problem (see above) 
is therefore necessary for the observability of the above effects.  
In the best GaAs/AlGaAs DQW systems $\Gamma \sim 0.25$meV~\cite{Eisen} and
the condition $eV > \Gamma$ can be easily satisfied.
\cite{Neil}.

As for the 1DES, the effects of disorder should be largely absent in the
consideration of carbon nanotubes and organic polymers themselves.  However
achieving a 2DES sufficiently close to a heterostructure surface to allow
tunneling into these systems is a technologically difficult problem and the
resulting 2DES is likely to have a larger $\Gamma$ than a fully optimized DQW
structure \cite{ritchie}.  For a surface gate defined 1DES in a DQW the remote
ionized impurities, random impurities and crystal faults could be strong
enough to pin its low lying excitations thereby destroying the LL behavior.
However, provided this does not happen, i.e. assuming that a LL phase in
quantum wires may exist {\it in principle}~\cite{fn3}, we expect the disorder
to effectively renormalize the characteristic LL parameters, most notably the
spin and charge density wave velocities.  Similarly, variations in the
thickness of the tunnel barrier can in principle have a large effect on
tunneling rates and therefore the clarity of any measured signal.  
At any rate, neither the presence of remote impurities nor tunnel barrier
variations have prevented experiments in high mobility DQW systems from
clearly resolving structures of the spectral functions of
quasi-one-dimensional systems \cite{kardynal97}.

Summarizing, we have proposed an experiment which should allow the
detection of Luttinger liquid behavior in a 1DES by detecting
magnetotunneling between the 1DES and a parallel 2DES. We have shown
how the parameters characterizing a LL, the ratio of spin and charge
velocity, can be determined from the voltage and/or magnetic field
dependence of the tunneling conductance. It was argued that,
notwithstanding the presence of thermal and disorder smearing effects,
the experiment should be feasible by means of today's technology.

{\it Acknowledgements:} We are grateful to I.~Aleiner and C. J. B Ford
for instructive discussions. FWJH acknowledges the financial
support of the European Community (contract ERB-CHBI-CT941764);
AA and FWJH acknowledge the support of the Deutsche Forschungsgemeinschaft
through SFB 237. CHWB
and AJS acknowledge the support of the EPSRC and the Royal Society
respectively.

\end{multicols}

\begin{references}
\bibitem{fn2}
$A_{2D}$ is defined as the $1D$ Fourier transform of
the $2D$ spectral function evaluated on the line $(x,0)$ and involves an
integration over all momenta in $y$-direction. As a consequence
$A_{2D}(q,\omega)$ contains spectral weight for small momenta $|v_F
q|<\omega$. The physical meaning of this `shadow region'  is that
states propagating between two points on the line $(x,0)$ can `detour'
through the extended plane, thereby propagating effectively slower
than with the Fermi velocity. We use a non-interacting form to
caption the singular structure of $A_{2D}$ due to quasiparticles
at $T=0$.
\bibitem{fn1}
For repulsive interactions, $v_\rho>v_\sigma$ holds in general. The
  first inequality $v_F>v_\rho$ is a condition which can be set 
  experimentally by varying the 2DES density.
\bibitem{fn3} In any given realization of this experiment it would be possible
  to determine whether disorder in the 1D system is disrupting the Luttinger
  liquid state by monitoring the conductance of the wire and ensuring that it
  is $2e^2/h$.
\end{references}
\end{document}